\documentclass[prl,reprint,longbibliography,superscriptaddress]{revtex4-2}

\usepackage{graphicx}
\usepackage{amssymb,amsmath,amsthm,bm}
\usepackage{physics}
\usepackage{lipsum}
\usepackage{float}
\usepackage{xcolor}
\definecolor{myred}{RGB}{216,18,116} 
\usepackage[export]{adjustbox}

\usepackage{multirow} 
\usepackage{verbatim}
\usepackage{booktabs}
\usepackage{siunitx}
\usepackage{gensymb}
\usepackage{makecell}
\usepackage{balance}
\usepackage[colorlinks=true,linkcolor=myred,citecolor=myred, pdfauthor={ },pdftitle={ },pdfsubject={ },pdfkeywords={ }]{hyperref}

\begin{document}

\title{Experimental realization and synchronization of a quantum van der Pol oscillator}

\author{Yi Li}
\thanks{These authors contribute equally to this work.}
\affiliation{CAS Key Laboratory of Microscale Magnetic Resonance and School of Physical Sciences, University of Science and Technology of China, Hefei 230026, China}
\affiliation{Anhui Province Key Laboratory of Scientific Instrument Development and Application, University of Science and Technology of China, Hefei 230026, China}
\affiliation{Hefei National Laboratory, University of Science and Technology of China, Hefei 230088, China}

\author{Zihan Xie}
\thanks{These authors contribute equally to this work.}
\affiliation{CAS Key Laboratory of Microscale Magnetic Resonance and School of Physical Sciences, University of Science and Technology of China, Hefei 230026, China}
\affiliation{Anhui Province Key Laboratory of Scientific Instrument Development and Application, University of Science and Technology of China, Hefei 230026, China}
\affiliation{Hefei National Laboratory, University of Science and Technology of China, Hefei 230088, China}

\author{Xiaodong Yang}
\email{yangxd@szu.edu.cn}
\affiliation{Institute of Quantum Precision Measurement, State Key Laboratory of Radio Frequency Heterogeneous Integration, College of Physics and Optoelectronic Engineering, Shenzhen University, Shenzhen 518060, China}
\affiliation{Quantum Science Center of Guangdong-Hong Kong-Macao Greater Bay Area (Guangdong), Shenzhen 518045, China}

\author{Yue Li}
\affiliation{CAS Key Laboratory of Microscale Magnetic Resonance and School of Physical Sciences, University of Science and Technology of China, Hefei 230026, China}
\affiliation{Anhui Province Key Laboratory of Scientific Instrument Development and Application, University of Science and Technology of China, Hefei 230026, China}

\author{Xingyu Zhao}
\affiliation{CAS Key Laboratory of Microscale Magnetic Resonance and School of Physical Sciences, University of Science and Technology of China, Hefei 230026, China}
\affiliation{Anhui Province Key Laboratory of Scientific Instrument Development and Application, University of Science and Technology of China, Hefei 230026, China}
\affiliation{Hefei National Laboratory, University of Science and Technology of China, Hefei 230088, China}

\author{Xu Cheng}
\affiliation{CAS Key Laboratory of Microscale Magnetic Resonance and School of Physical Sciences, University of Science and Technology of China, Hefei 230026, China}
\affiliation{Anhui Province Key Laboratory of Scientific Instrument Development and Application, University of Science and Technology of China, Hefei 230026, China}
\affiliation{Hefei National Laboratory, University of Science and Technology of China, Hefei 230088, China}

\author{Xinhua Peng}
\affiliation{CAS Key Laboratory of Microscale Magnetic Resonance and School of Physical Sciences, University of Science and Technology of China, Hefei 230026, China}
\affiliation{Anhui Province Key Laboratory of Scientific Instrument Development and Application, University of Science and Technology of China, Hefei 230026, China}
\affiliation{Hefei National Laboratory, University of Science and Technology of China, Hefei 230088, China}

\author{Jun Li}
\email{lijunquantum@szu.edu.cn}
\affiliation{Institute of Quantum Precision Measurement, State Key Laboratory of Radio Frequency Heterogeneous Integration, College of Physics and Optoelectronic Engineering, Shenzhen University, Shenzhen 518060, China}
\affiliation{Quantum Science Center of Guangdong-Hong Kong-Macao Greater Bay Area (Guangdong), Shenzhen 518045, China}

\author{Eric Lutz}
\email{eric.lutz@itp1.uni-stuttgart.de}
\affiliation{Institute for Theoretical Physics I, University of Stuttgart, D-70550 Stuttgart, Germany}

\author{Yiheng Lin}
\email{yiheng@ustc.edu.cn}
\affiliation{CAS Key Laboratory of Microscale Magnetic Resonance and School of Physical Sciences, University of Science and Technology of China, Hefei 230026, China}
\affiliation{Anhui Province Key Laboratory of Scientific Instrument Development and Application, University of Science and Technology of China, Hefei 230026, China}
\affiliation{Hefei National Laboratory, University of Science and Technology of China, Hefei 230088, China}

\author{Jiangfeng Du} 
\email{djf@ustc.edu.cn}
\affiliation{CAS Key Laboratory of Microscale Magnetic Resonance and School of Physical Sciences, University of Science and Technology of China, Hefei 230026, China}
\affiliation{Anhui Province Key Laboratory of Scientific Instrument Development and Application, University of Science and Technology of China, Hefei 230026, China}
\affiliation{Hefei National Laboratory, University of Science and Technology of China, Hefei 230088, China}
\affiliation{Institute of Quantum Sensing and School of Physics, Zhejiang University, Hangzhou 310027, China}

\begin{abstract}
Classical self-sustained oscillators, that  generate periodic motion without periodic external forcing, are ubiquitous in science and technology. The realization of nonclassical self-oscillators is an important goal of quantum physics. We here present the experimental implementation of a quantum van der Pol oscillator, a paradigmatic autonomous quantum driven-dissipative system with nonlinear damping,  using a single trapped atom. We demonstrate the existence of a quantum limit cycle in phase space in the absence of a drive and the occurrence of quantum synchronization when the nonlinear oscillator is externally driven. We additionally  show that synchronization can be enhanced with the help  of squeezing perpendicular to the direction of the drive and, counterintuitively, linear dissipation. We also observe the bifurcation to a bistable phase-space distribution for large squeezing. Our results pave the way for the exploration of self-sustained quantum oscillators and their application to \mbox{quantum technology.} 
\end{abstract}

\maketitle

The van der Pol  oscillator is a prototypical self-sustained oscillator with nonlinear friction \cite{jen13}. Due to its versatility, it has played a central role in the theory of nonlinear oscillations \cite{naf79,guc83}, the exploration of chaotic dynamics  \cite{str94,tho02} and the study  of synchronization \cite{pik01,osi07}. The nonlinear damping is such that the van der Pol oscillator is supplied with energy  for small amplitudes, while energy is dissipated for large amplitudes. Self-oscillations hence occur in the absence of any external force, and the system asymptotically tends to a limit cycle \cite{naf79,guc83}. Self-sustained oscillators differ from ordinary nonlinear systems, since they cannot be analyzed using quasilinear methods. Periodic forcing of the van der Pol oscillator leads to chaotic attractors owing to the nontrivial competition between external cyclic driving and internal autonomous oscillations \cite{str94,tho02}.  Its limit cycle can additionally be entrained by a weak periodic signal, inducing   synchronized oscillations \cite{pik01,osi07}. 

\begin{figure}
  \centering
  \includegraphics[width=0.48\textwidth]{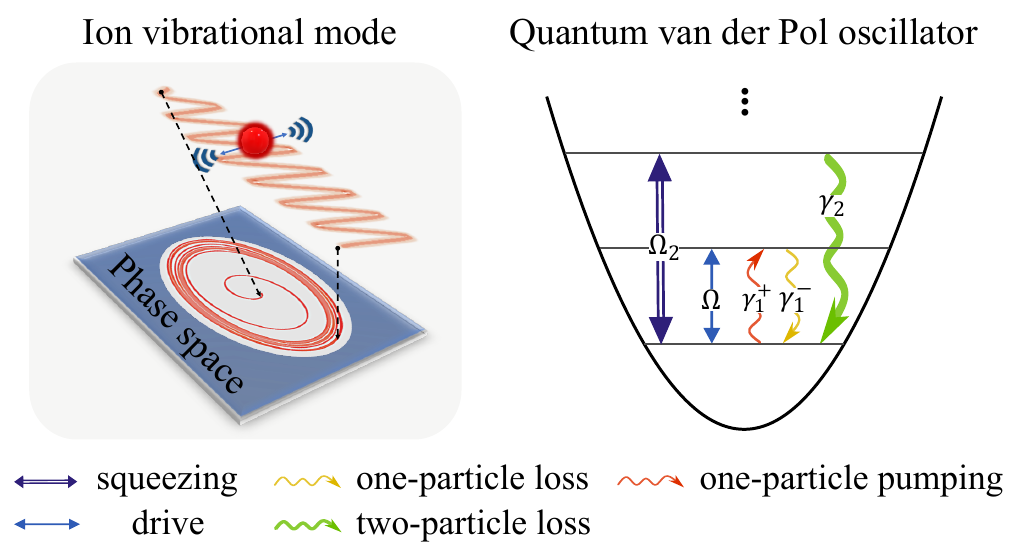}
  \caption{Schematics of the trapped-ion quantum van der Pol oscillator. An ion  oscillates around its equilibrium position, forming a vibrational mode subjected to  linear and nonlinear dissipation that yields  a limit cycle in phase space. The driven-dissipative dynamics is implemented through multiple simultaneous processes: coherent driving with strength $\Omega$, single-particle pumping at rate $\gamma_1^+$, single-particle loss at rate $\gamma_1^-$ (linear damping),   two-particle loss at rate $\gamma_2$ (nonlinear damping), and, optionally, squeezing with strength $\Omega_2$.}  \label{scheme}
\end{figure}

\begin{figure*}[ht]
    \centering
    \includegraphics[width=0.95\textwidth]{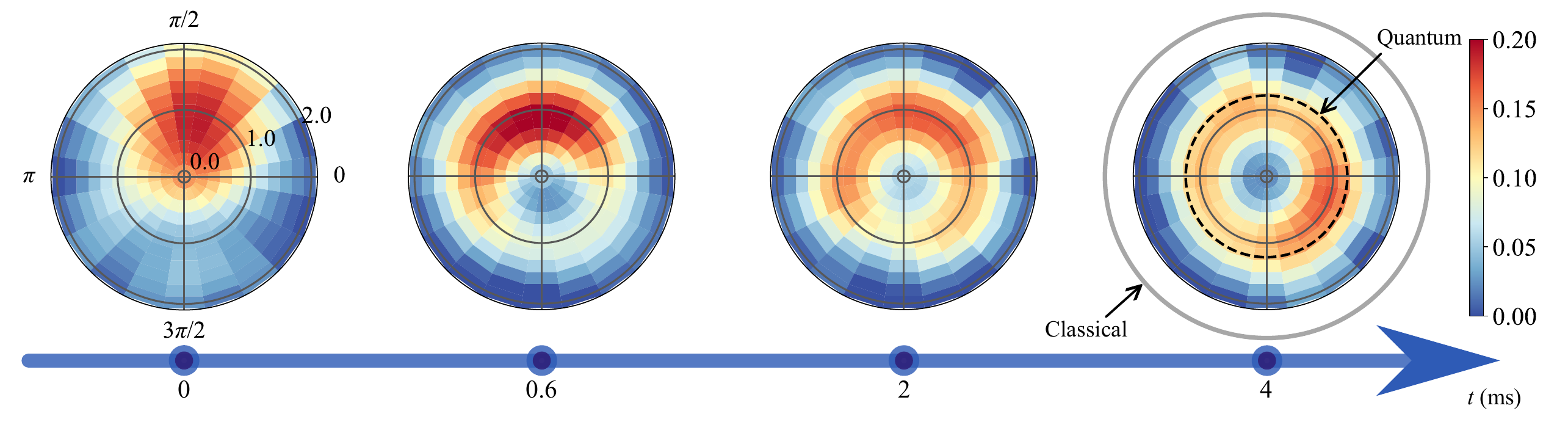}
    \caption{Quantum limit cycle of the undriven quantum van der Pol oscillator. Starting from a displaced thermal state ($\bar{n}=1.5, \alpha=1$),  the reconstructed Wigner function $W(r,\phi)$  evolves into a ring-shape steady-state phase-space distribution, associated with a quantum limit cycle. The radius of the quantum limit cycle (black dashed circle) is a factor two smaller than the radius of the corresponding classical limit cycle (gray solid). The dissipative ratio is  $\gamma_2/(\gamma_1^+ - \gamma_1^- )= 0.56$.}
    \label{Fig2}
\end{figure*}

Theoretical research has recently focused on the quantum version of the van der Pol oscillator \cite{lee13,wal14,lee14a,wal15,ame15,lor16,wei17,son18,sca19,dut19,mok20,aro21,kat21,cab21}, a canonical nonlinear model of driven-dissipative  open  quantum systems \cite{mul12,rot15,sie16}. The tunable interplay of coherent drive and incoherent dissipation in these systems generates complex nonequilibrium evolution and nontrivial steady states \cite{mul12,rot15,sie16}, which make them interesting for quantum technological applications \cite{ver09}. The presence of nonlinearities in general and of self-sustained oscillations in particular further increase the complexity of the dynamics \cite{sca19}. The properties of quantum and classical  van der Pol oscillators usually strongly differ due to the effects of quantum fluctuations and  quantum coherence \cite{lee13,wal14,lee14a,wal15,ame15,lor16,wei17,son18,sca19,dut19,mok20,aro21,kat21,cab21}. Quantum noise is thus expected to hinder quantum synchronization \cite{wal14}, while squeezing \cite{son18} and dissipation \cite{mok20} are predicted to enhance it. A squeezed drive should also lead to quantum metastability  associated with large timescale separations in the dynamics \cite{cab21}. Even though the van der Pol oscillator \cite{pol26} has been introduced the same year as Schr\"odinger's equation \cite{sch26}, its quantum implementation  has remained elusive.

We here experimentally realize a quantum van der Pol oscillator using a single Ca$^+$ ion in a Paul trap \cite{lei03} using reservoir engineering techniques \cite{har22}. We analyze both the undriven and the forced nonlinear quantum oscillator, as well as the influence of squeezing. In the undriven case, we observe the approach to a quantum limit cycle in phase space by reconstructing \cite{PhysRevLett.125.043602} the evolution of the Wigner function \cite{hil84}. We further demonstrate the occurrence of stable phase synchronization to an external drive \cite{lee13}, and determine the corresponding Arnold tongue diagram \cite{mok20}. We also find that synchronization is enhanced by linear dissipation in the deep quantum regime, a behavior that is absent in the classical and quantum domains \cite{mok20}. Finally, we show that squeezing  applied perpendicularly to the drive in phase space induces a bifurcation of the steady-state Wigner function, and that it additionally increases \mbox{phase synchronization \cite{son18}.}

\begin{figure*}[]
    \includegraphics[width=1\textwidth]{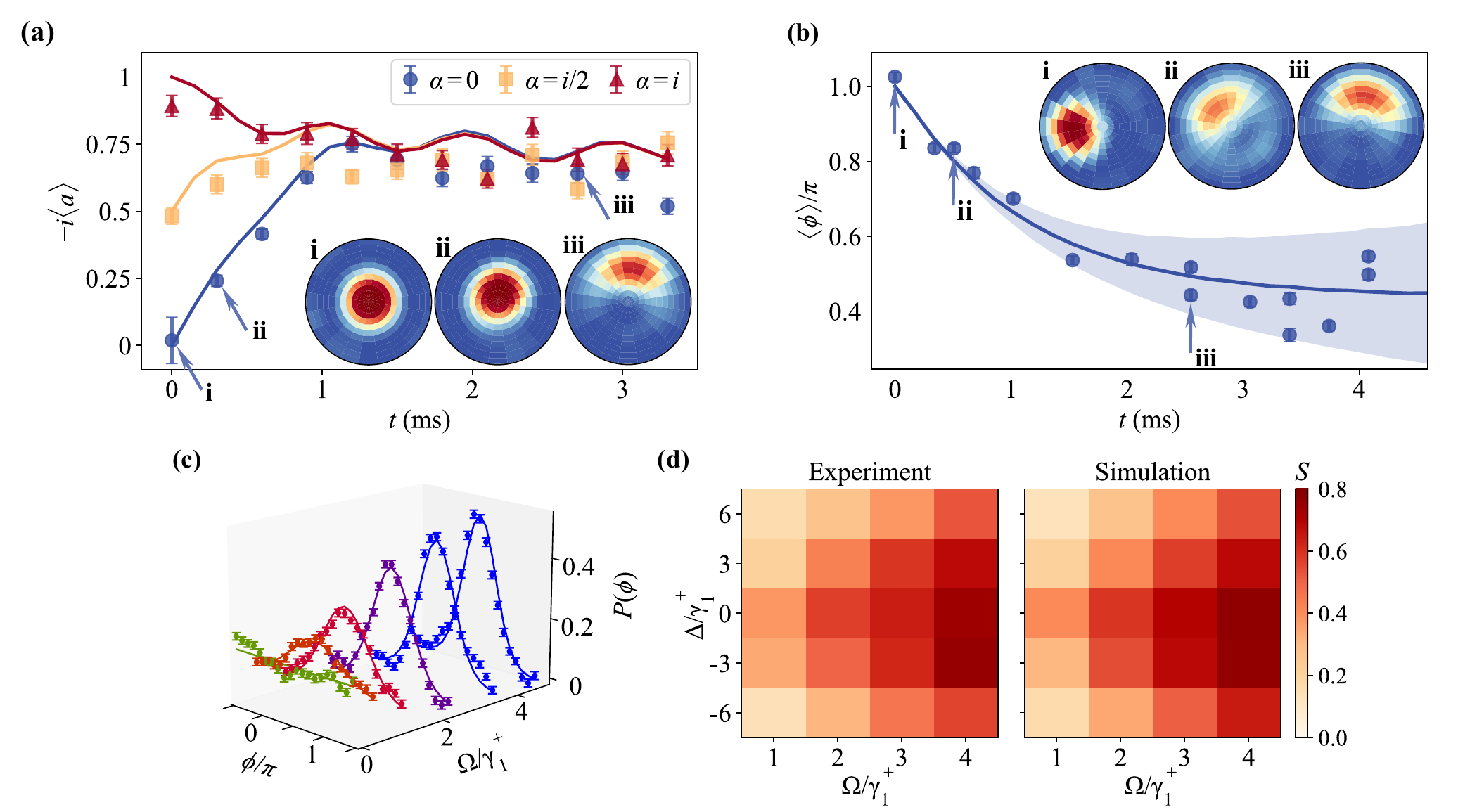}
    \caption{Phase synchronization of the driven quantum van der Pol oscillator. \textbf{(a)} The mean  amplitude $\langle a\rangle$  for different initial coherent states ($\alpha = 0, i/2, i$), with damping and driving parameters  $\gamma_2/\gamma_1^+ = 5.2$ and $\Omega/\gamma_1^+ = 3.5$, converges to the same steady oscillations, demonstrating entrainment between oscillator and  drive. Insets (i-iii) show the evolution of the Wigner function $W(r, \phi)$, starting from the vacuum state.
    \textbf{(b)} The  mean phase $\langle \phi\rangle$ for an initial coherent state $\alpha=-1$, with $\gamma_2/\gamma_1^+ = 5.7$ and $\Omega/\gamma_1^+ = 4.7$, locks to a constant value that slightly differs from the driving direction because of off-resonant effects. The shaded blue area indicates phase fluctuations due to experimental imperfections. Insets (i-iii) depict the evolution of the Wigner function, starting from this coherent state.  \textbf{(c)} The steady-state phase distribution $P(\phi)$ is a Gaussian centered on the driving phase, whose width decreases with increasing driving strength.      Parameters are $\gamma_2/\gamma_1^+ = 5.7$ and $\Omega/\gamma_1^+ = \{0, 0.47, 1.2, 2.4, 3.6, 4.7\}$. 
    \textbf{(d)} Arnold  tongue for the mean resultant length $S$ as a function of detuning and  strength of the driving, defining the synchronization region. The dissipative rate is $\gamma_2/\gamma_1^+ = 4.7$.}
    \label{Fig3}
\end{figure*}


{\emph{Experimental system}. We experimentally implement a quantum van der Pol oscillator whose dynamics can be described by the  master equation (in units of $\hbar$) \cite{lee13,wal14,lee14a,wal15,ame15,lor16,wei17,son18,sca19,dut19,mok20,aro21,kat21,cab21}
\begin{equation}\label{vdp}
	\dot \rho= - i [H, \rho] + \gamma_1^+ \mathcal{D}[a^\dag]\rho + \gamma_1^- \mathcal{D}[a]\rho  + \gamma_2 \mathcal{D}[a^2]\rho,
\end{equation}
where $H=-\Delta a^\dag a +i\Omega(a^\dag-a) +i\Omega_2/2 ( a^{\dag2} e^{2i\theta}-a^{ 2}e^{-2i\theta} )$ is the coherent Hamiltonian (with detuning $\Delta$), including the displacement drive (with strength $\Omega$) and squeezing drive in the rotating frame (with amplitude $\Omega_2$); the parameter $\theta$ is the relative phase between the two drives. The operators $a$ and $a^\dagger$ are the usual ladder operators of the harmonic oscillator, and the Lindblad dissipators  are given by $\mathcal{D}[O]\rho=O\rho O^\dag -\{O^\dag O,\rho\}/2$.
The two coefficients $\gamma_1^+$ and  $\gamma_1^-$  respectively represent the one-particle pumping rate and the  one-particle loss rate (associated with linear damping), while $\gamma_2$ is the two-particle loss rate (corresponding to nonlinear damping) (Fig.~\ref{scheme}).
Equation \eqref{vdp} differs from recent ion-trap realizations of self-excited phonon lasers \cite{vahala2009phonon,PhysRevLett.131.043605} through the presence of nonlinear friction. It exhibits distinct regimes depending on the relative dissipation rates:  semiclassical  for $\gamma_2/\gamma_1^+\lesssim0.1$,   quantum for $\gamma_2/\gamma_1^+\sim 1$, and  deep quantum for $\gamma_2/\gamma_1^+\gtrsim 10$ \cite{mok20}.  Our study primarily focuses on the latter two, where the phonon population remains low and the system displays nonclassical properties.

The experiment uses the axial motion of a single $\mathrm{^{40}Ca^+}$ trapped ion as the harmonic oscillator with a trapping frequency $\omega_m \approx (2\pi) \times 1.1 ~\mathrm{MHz}$. The internal spin states are defined as $\ket{\downarrow} = |S_{1/2}, m_J=+1/2\rangle$, $\ket{\uparrow} = |D_{5/2}, m_J=+1/2\rangle$, and $\ket{\mathrm{aux}} = |D_{5/2}, m_J=+5/2\rangle$ to assist laser induced spin-motion couplings with a narrow-linewidth 729 nm laser along the axial direction. 
The displacement drive in Eq.~\eqref{vdp} is achieved through a radio-frequency signal at frequency $\omega = \omega_m + \Delta$ applied to one of the trap electrodes~\cite{doi:10.1126/science.aaw2884}.  Linear and nonlinear damping processes are further realized by  reservoir engineering methods~\cite{har22}, coupling the oscillator to the spin such that  $\ket{\downarrow,n}\rightarrow\ket{\mathrm{aux},n\pm1}$ and $\ket{\downarrow,n}\rightarrow\ket{\mathrm{aux},n-2}$, with the spin optically pumped back to $\ket{\downarrow}$, effectively forming  the motional dissipation with respective strengths of $\gamma_1^\pm$ and $\gamma_2$ (Supplementary Information). We additionally  implement squeezing with controllable strength by  applying two pairs of laser tones off-resonantly driving the spin-motion joint transition~\cite{2403.05471}. After Doppler and ground state cooling of the motion \cite{lei03}, the displacement drive is then applied continuously, while other coupling and pumping lasers are applied sequentially via a first-order Trotterization approach  \cite{PhysRevLett.127.020504}.
Eventually, we investigate the  properties of the quantum van der Pol oscillator via tomographic state   reconstruction  that maps the oscillator's motional information onto spin populations, and thus directly measures the Wigner function in  polar coordinates $W(r, \phi)$~\cite{PhysRevLett.125.043602}. 

\begin{figure*}[htbp]
\centering
\includegraphics[width=1\textwidth]{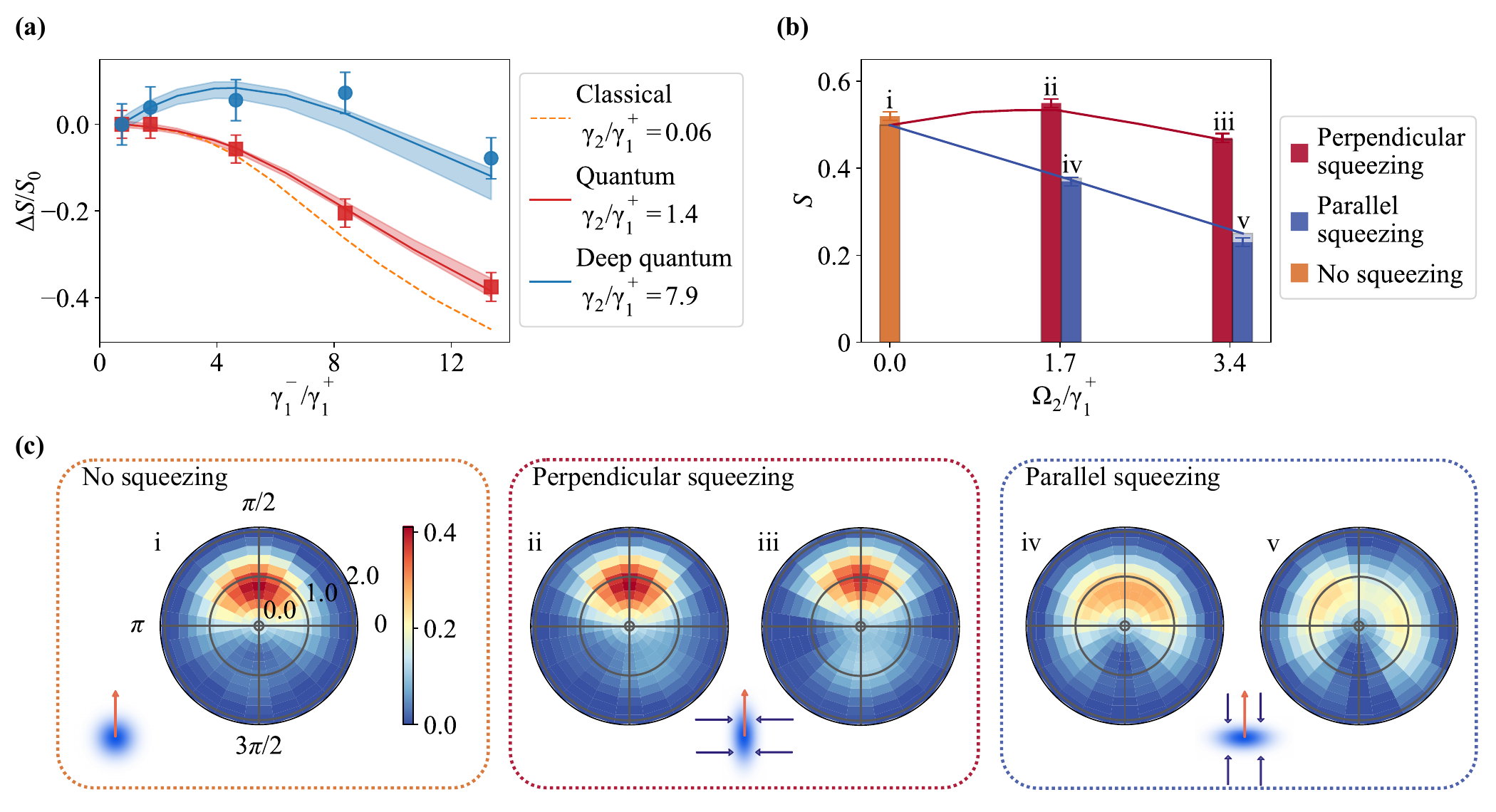}
    \caption{Quantum synchronization enhancement under dissipation and squeezing. \textbf{(a)} Moderate single-phonon dissipation increases synchronization in the deep quantum regime  ($\gamma_2 / \gamma_1^+=7.9$), but not in the quantum ($\gamma_2/\gamma_1^+=1.4$) and semiclassical ($\gamma_2/\gamma_1^+=0.06$) domains.  The external drive is resonant with the oscillator frequency, with a strength of $\Omega/\gamma_1^+=3$. 
    \textbf{(b)} Moderate squeezing perpendicular  to the drive ($\theta=\pi/2$) enhances synchronization by suppressing  quantum fluctuations, whereas squeezing parallel to the drive ($\theta=0$) decreases it; parameters are $\gamma_2/\gamma_1^+=4.4$ and $\Omega/\gamma_1^+=1.2$.  
    \textbf{(c)} Corresponding Wigner functions to (c): (i) no squeezing $\Omega_2=0$ serves as a reference, while (ii, iv) and (iii, v) correspond to squeezing strengths of $\Omega_2 / \gamma_1^+=1.7$ and $\Omega_2 / \gamma_1^+=3.4$, respectively. In the schematics, blue arrows indicate the squeezing direction, whereas orange arrows represent the external drive direction. Strong parallel squeezing leads to a bifurcation to a bistable Wigner function (v).}
\label{Fig4}
\end{figure*}

{\emph{Quantum limit cycle and synchronization.} Let us begin by analyzing the dynamics of the undriven van der Pol oscillator, where $\Delta=\Omega=\Omega_2=0$. To that end, we prepare the oscillator in a displaced thermal state \cite{jeo06}, that is, a thermal state with mean phonon number $\bar{n} = 1.5$ displaced by a coherent state with amplitude $\alpha = 1.0$, and measure the  Wigner function $W(r, \phi)$ at times 0, 600, 2000, and 4000 $\mu$s for  $\gamma_2/(\gamma_1^+ - \gamma_1^- )= 0.56$ (Fig.~\ref{Fig2}), with $\gamma_1^{\{+,-\}}=\{2.06~ \rm{kHz}, 0.09~\rm{kHz}\}$. We observe that the initially localized state first relaxes to a lower amplitude before spreading in phase space to form the typical ring shape of a limit cycle \cite{lee13}. This establishes the autonomous nature of the oscillations.  We further note that the radius of the quantum ring (black dashed) is a factor two smaller than the  radius of the corresponding classical limit cycle $r_c= 2({\gamma_1^+ - \gamma_1^- }/{\gamma_2})^{0.5}=2.6$ (gray solid) \cite{jen13}, highlighting its nonclassical nature \cite{aro21}.

We next switch on the resonant coherent drive with amplitude $\Omega$ according to Eq.~\eqref{vdp}, oriented along the $\pi/2$ direction  ($\Delta=\Omega_2=0$). The temporal evolution of the average amplitude $\langle a \rangle$ of the oscillator is measured for various initial conditions,  $\alpha = 0, i/2, i$, for $\gamma_2/\gamma_1^+ = 5.2$ and $\Omega/\gamma_1^+ = 3.5$ (Fig.~\ref{Fig3}a).  For all these initial conditions, the mean amplitude $\langle a \rangle$ converges to the same steady oscillations as the van der Pol oscillator adjusts itself to the external drive, indicating entrainment between the two  \cite{jen13}. The inset further shows that the drive breaks the radial symmetry of the undriven limit cycle, aligning the oscillator's phase-space distribution with the drive axis. Additional insight is obtained by examining the phase dynamics of an initial coherent state   with a phase offset relative to the drive $\alpha=-1$ (Fig.~\ref{Fig3}b): the mean phase $\langle \phi \rangle$  locks to a constant value close to $\pi/2$ (Fig.~\ref{Fig3}b), a hallmark of phase synchronization   \cite{lee13}. The shaded blue area accounts for experimental imperfections, including oscillator frequency and laser frequency fluctuations.  Experimental results align closely with numerical simulations (solid lines), with minor deviations (e.g., a slight phase misalignment) attributable to off-resonance effects caused by the sideband transitions (Supplementary Information). Figure~\ref{Fig3}c moreover reveals that the phase distribution, $P(\phi)=\int W(r,\phi)rdr$, adopts a Gaussian profile, centered at the drive phase, that becomes narrower with increasing driving strength $\Omega/\gamma_1^+$ ($\gamma_2/\gamma_1^+ = 5.7$).

 To quantify phase synchronization, we use the mean resultant length  of a circular distribution, $S=|\langle e^{i \phi}\rangle |= \left|\int_\phi e^{i\phi} P(\phi) d\phi\right|$ \cite{mok20}. It takes the value 0 for an unsynchronized state and the value 1 for a perfectly synchronized state,  which corresponds to a delta  Wigner function. 
We examine the  synchronization parameter $S$  as a function of driving strength $\Omega$ and detuning $\Delta$, while keeping the dissipative rate fixed at $\gamma_2/\gamma_1^+ = 4.7$ (Fig.~\ref{Fig3}d). We find that synchronization  increases with the driving strength and decreases with the detuning, as expected. We recognize a structure which is reminiscent of an Arnold tongue which defines the synchronized domain of classical synchronization phenomena \cite{pik01,osi07}. Figure~\ref{Fig3}d shows that phase locking of the quantum van der Pol oscillator is a robust phenomenon that appears in a large region of parameter space. We again have good agreement with numerical simulations. As before, there is a slight asymmetry in the diagram due to experiment imperfections.

{\emph{Quantum  dissipation boost and squeezing.}
Quantum noise is usually  expected to be detrimental for synchronization as it provides an additional source of phase diffusion \cite{wal14}. It is therefore essential to identify quantum mechanisms that allow one to boost synchronization in the quantum realm.  We first investigate the influence of single-phonon dissipation by increasing the parameter $\gamma_1^-$ in three different regimes defined by the strength of the nonlinear damping \cite{mok20}: $\gamma_2/\gamma_1^+ = 7.9$ (deep quantum regime), $\gamma_2/\gamma_1^+ = 1.4$ (quantum regime) and  $\gamma_2/\gamma_1^+ = 0.06$ (semiclassical regime). The external drive is  resonant with the oscillator's frequency, with a strength of $\Omega/\gamma_1^+=3$. Values of $\gamma_1^-\sim \{\gamma_1^+, \Omega\}$ are needed to avoid strong decoherence. We observe in Fig.~\ref{Fig4}a that  increasing single-phonon dissipation enhances synchronization ($\Delta S>0$) in the deep quantum regime (blue dots), but reduces it in the quantum (red squares) and semiclassical (yellow line) regimes; the small $\gamma_2$ domain is not accessible experimentally and hence simulated. This dramatic difference can be explained by noting that the decoherence rate grows with the phonon number \cite{mok20}, meaning that higher levels decohere faster. By driving the oscillator to lower levels, single-phonon dissipation thus initially helps  increase the transition rate between the two lowest levels, that is, build up  coherence  between  $|0\rangle$ and $|1\rangle$, which in turn, augments synchronization. Larger single-phonon dissipation suppresses quantum synchronization ($\Delta S<0$) in all  regimes, as dissipation-induced decoherence becomes dominant.

Another strategy to counteract the negative influence of quantum noise, and augment synchronization, is squeezing \cite{son18}. Squeezing is indeed able to suppress fluctuations in one direction, at the expense of the orthogonal direction. Figure \ref{Fig4}c shows the reconstructed Wigner function for three values of the squeezing strength ($\Omega_2/\gamma_1^+=\{0, 2.435, 4.870\}$), when the squeezing direction is   either parallel ($\theta=\pi/2$) or perpendicular ($\theta=0$) to the direction of the displacement drive; the dissipative parameter is here $\gamma_2 / \gamma_1^+=4.4$ and the resonant driving strength is $\Omega/\gamma_1^+=1.2$. 
The shape of the Wigner function results from the competition between the displacement drive and the squeezing drive. For small squeezing strengths, the Wigner function remains primarily aligned with the displacement direction. However, as the squeezing amplitude increases, and exceeds the displacement strength, the Wigner function splits, signaling  the bifurcation from a monostable distribution to a bistable distribution \cite{cab21}.  
 Small squeezing perpendicular to the drive ($\theta=0$) reduces the phase variance, thus enhancing synchronization, as seen in the increase of the mean resultant length $S$ in Fig.~\ref{Fig4}c; this effect could be further improved by optimizing the alignment between the two drives. By contrast, small squeezing parallel to the drive ($\theta=\pi/2$) increases the phase variance, which hinders synchronization.  In both instances, a larger squeezing parameter, beyond the bifurcation to multistability, reduces the mean resultant length $S$, beyond the level achieved without squeezing. One may hence conclude that there is an optimal  parameter window in which squeezing has a positive effect on quantum synchronization, for instance shown in Fig.~\ref{Fig4}b.

{\emph{Conclusion.} Driven-dissipative quantum systems are significantly influenced by the dissipative part of the dynamics, rather than by their Hamiltonian, and therefore exhibit properties drastically different from those of closed systems at equilibrium. By engineering nonlinear dissipation, we have demonstrated the experimental realization of an autonomous  quantum van der Pol oscillator using a single trapped ion. Our results offer key insights into quantum limit cycles, phase synchronization, and the interplay between coherence, dissipation and squeezing. They thus provide a valuable platform to investigate the intricate dynamics of quantum self-sustained oscillators. They could, for example,  be harnessed for applications in quantum sensing \cite{dut19}, quantum computation and communication \cite{ver09}, or improve the control and manipulation of quantum states \cite{leg15}. Moreover, adding higher-order nonlinear drives could enable the exploration of exotic quantum phases \cite{lab23}, while extensions to networks of nonlinear oscillators could reveal  novel collective phenomena and dissipative phase transitions \cite{dav18}. These systems thus hold immense potential for advancing quantum technologies and deepening our understanding of nonequilibrium quantum dynamics. 

We thank Yangchao Shen for helpful discussion. The USTC team acknowledges support from the National Natural Science Foundation of China (Grant No.~92165206) and the Innovation Program for Quantum Science and Technology (Grant No.~2021ZD0301603). Jun Li acknowledges support from the National Natural Science Foundation of China (Grant No.~12441502 and No.~12122506), the Shenzhen Science and Technology Program (RCYX20200714114522109). Xiaodong Yang acknowledges support from the National Natural Science Foundation of China (Grant No.~12204230). Eric Lutz is supported by the German Science Foundation DFG (Grant No.~FOR 2724).

\onecolumngrid   
\newpage

\begin{center}
    \textbf{Supplementary Materials for:\\``Experimental realization and synchronization of a quantum van der Pol oscillator”}
\end{center}

\twocolumngrid   

\appendix
\setcounter{figure}{0}
\renewcommand{\thefigure}{S\arabic{figure}}  
\setcounter{table}{0}
\renewcommand{\thetable}{S\arabic{table}}    
\setcounter{equation}{0}
\renewcommand{\theequation}{S\arabic{equation}} 

\renewcommand{\theHfigure}{S.\arabic{figure}} 
\renewcommand{\theHtable}{S.\arabic{table}}
\renewcommand{\theHequation}{S.\arabic{equation}}


\maketitle

\section{Experimental scheme}
We demonstrate the quantum van der Pol oscillator through the Trotterization approach \cite{PhysRevLett.127.020504}: the target evolution is divided into $N$ steps, each consisting of coherent operations $e^{iH_j t/N}$ and dissipative operations $e^{K_j t/N}$, implemented alternately. The total operations can then mimic the original process: $\lim_{N\rightarrow \infty} \left (\Pi_j e^{iH_j t/N} e^{K_j t/N}\right )^N=e^{\sum_j(iH_j+K_j)t}$. The schematic of the circuit is shown in Fig.~\ref{Fig. S1}.

For the coherent part, a radio-frequency (RF) signal with frequency $\omega = \omega_m + \delta$ is applied to 
 one of the electrodes to generate the displacement driving \cite{doi:10.1126/science.aaw2884}. This is the only component always applied during the sequence, other parts are implemented sequentially. We introduce phonon squeezing along the controllable direction and strength by laser driving two non-commuting spin-dependent forces \cite{2403.05471}.

For the dissipative part, the internal spins are introduced as the dissipative channel to engineer the desired jump operators $\mathcal{D}[O]$ through sideband transitions and spin reset. Since the three dissipation channels, including $\mathcal{D}[a^\dag]$, $\mathcal{D}[a]$ and $\mathcal{D}[a^2]$, undergo the similar operation, we only take $\mathcal{D}[a]$ as an example. Considering the spin-phonon system whose initial state is $\rho\otimes\left|\downarrow\right\rangle\left\langle\downarrow\right|$, the red sideband Hamiltonian is
\begin{equation}
H_{rsb}=\Omega_{rsb}/2(\sigma^+a+\sigma^- a^\dagger),
\end{equation}
where $\Omega_{rsb}$ is the Rabi frequency of the first-order red sideband. After sideband transition time $\tau_{rsb}$, the density matrix for the oscillator is
\begin{equation}
\begin{aligned}
\rho(t+\tau_{rsb})&=Tr_{spin}[e^{-iH\tau_{rsb}}\rho\otimes\left|\downarrow\right\rangle\left\langle\downarrow\right|e^{iH\tau_{rsb}}]\\
&=\left\langle\uparrow\right|\rho_{tot}(t+\tau_{rsb})\left|\uparrow\right\rangle+\left\langle\downarrow\right|\rho_{tot}(t+\tau_{rsb})\left|\downarrow\right\rangle\\
&\equiv\rho_u+\rho_d.
\end{aligned}
\end{equation}
For small times $\tau_{rsb}$, we can expand the unitary operators as $e^{\pm iH\tau_{rsb}}=1\pm iH\tau_{rsb}+o(\tau_{rsb}^2)$ and obtain
\begin{equation}
\begin{aligned}
\rho_u&\approx\Gamma\tau_{rsb} a\rho(t)a^\dagger,\\
\rho_d&\approx\rho(t)-\frac{\Gamma}{2}\tau_{rsb}\left[a^\dagger a\rho(t)+\rho(t)a^\dagger a\right]+o(\tau_{rsb}^4),
\end{aligned}
\end{equation}
where $\Gamma=(\frac{\Omega_{rsb}}{2})^2\tau_{rsb}$. After the spin reset, the density matrix of the oscillator would become $\rho=\rho_u+\rho_d$, which undergoes the same trajectory as $\mathcal{D}[a]$ with effective pumping rate $\Gamma$.

\begin{figure}[t]
  \centering
  \includegraphics[width=0.97\columnwidth]{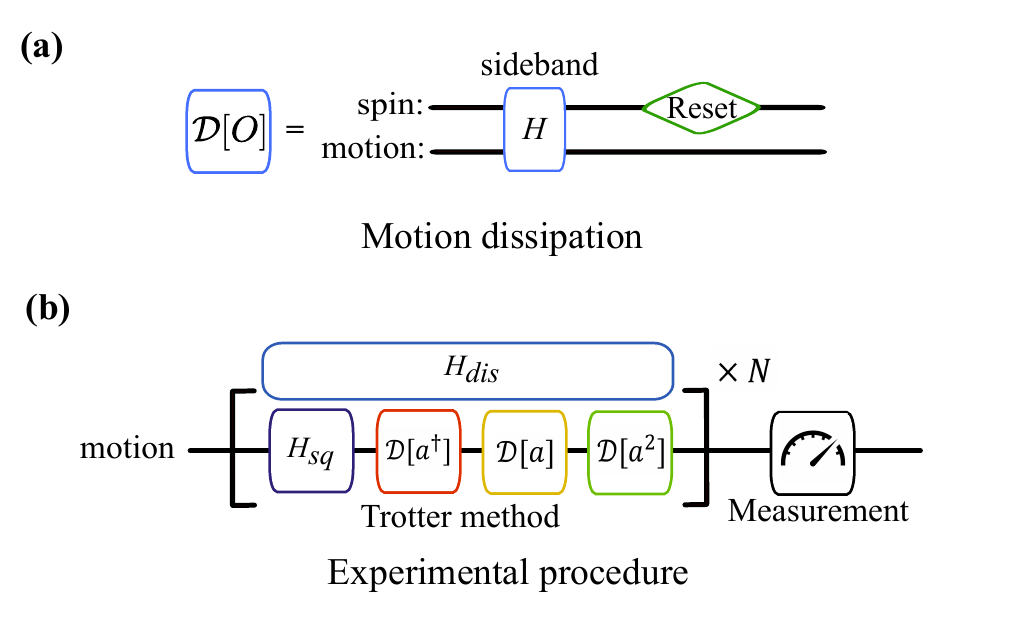}
  \caption{Schematic diagram of the experimental operation of the quantum van der Pol oscillator. \textbf{(a)} Motion dissipation, including one or two-particle pumping or loss, is achieved by the spin-motion coupling followed by a spin reset. \textbf{(b)} Part of the experimental procedure, including dynamical evolution and measurement. The squeezing and phonon dissipation are applied 
 by the Trotterization, while the external drive is applied continuously.}  \label{Fig. S1}
\end{figure}

In the experiment, we sequentially apply coherent and dissipative operations similar to above for respective duration $\tau$'s. The first-order Trotterization is applied as shown in Fig.~\ref{Fig. S1}, where each coherent or Lindblad operator is applied once for each cycle of duration $T$. As a result, the averaged strength of these drives in the Trotterization process needs to be rescaled with a factor $\tau/T$. Thus, the effective dissipative rate for $\mathcal{D}[a]$ is 
\begin{equation}
\gamma_{1}^-=\left(\frac{\Omega_{rsb}}{2}\right)^2\frac{\tau_{rsb}^2}{T}.
\end{equation}
For the other dissipative and coherent operators, we similarly obtain
\begin{equation}
\begin{aligned}
\gamma_{1}^+&=\left(\frac{\Omega_{bsb}}{2}\right)^2\frac{\tau_{bsb}^2}{T},\\
\gamma_{2}&=\left(\frac{\Omega_{2rsb}}{2}\right)^2\frac{\tau_{2rsb}^2}{T},\\
\Omega_{2}&=\frac{\Omega_{sq}\tau_{sq}}{T},
\end{aligned}
\end{equation}
where $\Omega_{bsb}, \Omega_{2rsb}, \Omega_{sq}$ represent the Rabi frequency of first-order blue sideband, second-order red sideband and squeezing respectively, $\tau_{bsb}, \tau_{2rsb}, \tau_{sq}$ are the corresponding pulse time.

\section{Experimental apparatus and procedure}
In this work, we use a single trapped $^{40}\rm{Ca}^+$ ion in a linear Paul trap \cite{lei03}. The quadrupole transition between $|S_{1/2}\rangle$ and $|D_{5/2}\rangle$ is used for coherent manipulations, which is driven by a narrow-linewidth 729~nm laser along the axial direction. The Lamb-Dicke parameter for the spin-motion coupling is $\eta=0.0925$. As the experiment is performed mostly in the (deep) quantum regime, the phonon is at most excited to 4, which means that the experiments can be well described by the Lamb-Dicke approximation~\cite{lei03}. The internal electronic state of the ion is initialized to $|\!\!\downarrow\rangle$ by the frequency-resolved 729 nm laser combined with 854 nm and 866 nm repump lasers. We use the 397~nm cycling transition between $|S_{1/2}\rangle$ and $|P_{1/2}\rangle$ along with 866 nm repump laser for fluorescence detection.

The axial mode is used to simulate the van der Pol oscillator. Its frequency is measured by ``tickle" spectroscopy, where we apply the radio frequency to one electrode and detect the resulting phonon population using the red sideband. We calibrate it every 30 minutes to ensure the long-term stability of the mode frequency within 30 Hz.

Before each experiment, the motional mode is initialized by Doppler cooling and sideband cooling. After sideband cooling, the mean phonon number of the mode can be cooled to less than 0.1. The subsequent experimental procedure involves phonon state preparation (if necessary), dynamic evolution and mode measurement. We prepare the coherent thermal state using an electronic displacement drive, combined with Trotterizations involving blue and red sidebands to simulate the heating process. During the dynamic evolution process, the external drive remains on (except for Fig.~2) throughout the entire evolution. The other operations are segmented. In each cycle, we sequentially apply squeezing (if necessary), one-particle pumping, one-particle loss (if necessary), two-particle loss, and spin state preparation. 
The state preparation is added to avoid the leakage to $|S_{1/2}, m_J=-1/2\rangle$. The phonon tomography is done right after the evolution. The whole procedure is shown in Fig.~\ref{Fig. S1}b and on Tab.~\ref{tab:1}. Key effective intensity parameters for the experiments are listed in Tab.~\ref{tab:3}. The pulse time and number of cycles are shown in Tab.~\ref{tab:4}.

\begin{table}
    \centering
    \begin{tabular}{l}
        \toprule
        Experimental procedure\\
        \midrule
         1. Doppler cooling and sideband cooling\\
         2. Spin state preparation\\
         3*. Prepare the initial state of phonon\\
         4*. Open the external drive\\
         5*. Squeezing drive\\
         6. One-particle pumping\\
         7*. One-particle loss\\
         8. Two-particle loss\\
         9. Spin state preparation\\
         10. Repeat Step 5-9 for N times\\
         11. Close the external drive\\
         12. Phonon measurement\\
         \bottomrule
    \end{tabular}
    \caption{The procedure of the van der Pol oscillator experiment. Steps marked with (*) are only implemented when required.}
    \label{tab:1}
\end{table}

\renewcommand{\arraystretch}{2.2}
\begin{table*}
    \centering
    \begin{tabular}{llllllll}
        \toprule
        Figure& $\gamma_1^+$ [kHz]& $\gamma_1^-$ [kHz]& heating rate [kHz]& $\gamma_2$ [kHz] &  $\Omega/2 \pi$ [Hz]& $\Omega_2/2\pi$ [Hz]&$\Delta/2\pi$ [Hz]\\
        \midrule
        Fig.2& 2.06& 0.09& 0.09& 1.11& 0& 0&0\\  
        Fig.3a& 0.28& 0.12& 0.12& 1.48& 160& 0&0\\ 
        Fig.3b& 0.23& 0.09& 0.09& 1.31& 173& 0&0\\ 
        Fig.3c& 0.23& 0.09& 0.09& 1.31& \makecell[l]{\{0, 17, 43, 87,\\ 130, 173\}}& 0&0\\ 
        Fig.3d& 0.28& 0.12& 0.12& 1.48& \makecell[l]{\{45, 91, 136, 181\}}& 0&\makecell[l]{\{-272, -136, 0,\\ 136, 272\}}\\ 
        Fig.4a quantum& 0.16& \makecell[l]{\{0.12,  0.27,  0.74,\\  1.33, 2.12\}}& 0.12& 0.22& 76& 0&0\\ 
        Fig.4a deep quantum& 0.16& \makecell[l]{\{0.12,  0.27,  0.74,  \\1.33, 2.12\}}& 0.12& 1.25& 76& 0&0\\
        Fig.4b-c& 0.23& 0.12& 0.12& 1.01& 43& \makecell[l]{\{0, 32($\perp$), 63($\perp$),\\ 32($\parallel$), 63($\parallel$)\} }&0\\
        \bottomrule \hline
    \end{tabular}
    \caption{Key parameters in the experiments.}
    \label{tab:3}
\end{table*}

\renewcommand{\arraystretch}{2.2}
\begin{table*}
    \centering
    \begin{tabular}{lllllllll}
        \toprule
        Figure& $\tau_{bsb}$ [$\mu s$]& $\tau_{rsb}$ [$\mu s$]& $\tau_{2rsb}$ [$\mu s$]& $\tau_{sq}$ [$\mu s$]&$\tau_{reset}$ [$\mu s$] & $\tau_{idle}$ [$\mu s$]& $T$ [$\mu s$]& $N$\\
        \midrule
        Fig.2& 40& 0& 150& 0& 10& 0&200& \makecell[l]{\{0, 3, 10, 20\}}\\  
        Fig.3a& 10& 0& 120& 0& 10& 10&150& \makecell[l]{\{0, 2, 4, ..., 22\}}\\ 
        Fig.3b& 10& 0& 150& 0& 10& 0&170&\makecell[l]{\{0, 2, 3, 4, 6, 9, 12, 15, 18, 20, 22, 24\}}\\ 
        Fig.3c& 10& 0& 150& 0 & 10& 0& 170&20\\ 
        Fig.3d& 10& 0& 120& 0 & 10&10& 150&48\\ 
        Fig.4a-b quantum& 5& 10& 50& 0 & 15& 80& 160& 37\\ 
        Fig.4a-b deep quantum& 5& 10& 120& 0& 15&10&160&37\\
        Fig.4c-d& 10& 0& 150& 35& 15&10& 220&20\\
        \bottomrule \hline
    \end{tabular}
    \caption{Pulse time in the experiments. $\tau_{rsb}$, $\tau_{bsb}, \tau_{2rsb}, \tau_{sq}$ are the corresponding pulse time for first-order red sideband, first-order blue sideband, second-order red sideband and squeezing respectively. $\tau_{reset}$ represents the total spin reset and initialization time in one cycle. $\tau_{idle}$ denotes the idle time in one cycle. $T$ is the total time for each cycle. $N$ is the number of cycles.}
    \label{tab:4}
\end{table*}

\begin{figure}[htbp]
  \centering
  \includegraphics[width=0.48\textwidth]{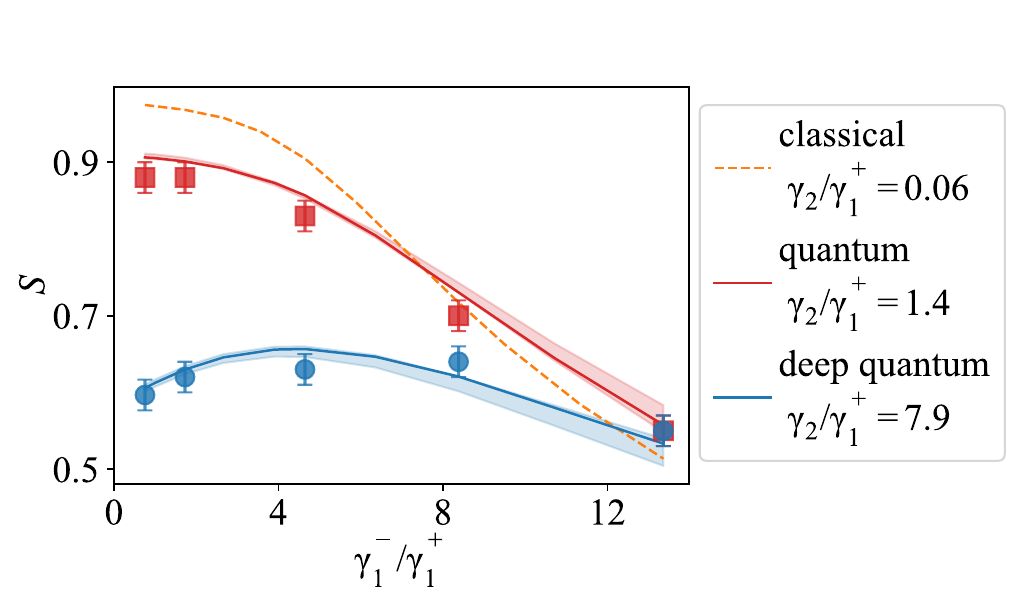}
  \caption{Mean resultant length ($S$) as a function of single-phonon dissipative rate. Synchronization is consistently stronger in the quantum regime compared to the deep quantum regime, as quantum noise is lower in the former.} 
  \label{Fig. S2}
\end{figure}

\section{Numerical Simulations and Experimental Imperfections}

We simulate the experimental sequence by modeling a two-level system coupled to a phonon mode truncated at 30 excitations. We use the full Hamiltonian instead of a simple linear expansion for the interaction between spin and motion. As will be discussed below, this gives rise to the motion phase shift caused by the off-resonant coupling of the sideband transitions. These simulations are performed using the QuTip toolbox~\cite{JOHANSSON20121760}. The simulation lines shown in the main text all consider the real experimental issues, including heating rate of phonon and full Hamiltonian of spin-motion couplings. The other imperfections are mainly due to the trapping frequency fluctuation, which is shown as the shadow region in the figures. 

Another key experimental issue is the unavoidable off-resonant coupling effect caused the second red sideband transition, which is used in conjunction with spin pumping to implement the two-phonon dissipator \(\mathcal{D}[a^2]\). Due to the small Lamb-Dicke parameter, additional coupling to the first red sideband and carrier cannot be neglected. A second-order Magnus expansion shows that this off-resonant interaction generates an effective phonon AC Stark shift, \((\eta \Omega_2^-)^2/(2\omega_z)\sigma_z a^\dagger a\), causing a slight shift in the steady-state phase relative to the external drive and minor oscillations during evolution. Nevertheless, these off-resonant effects have only a minimal impact on the mean resultant length compared to the ideal case. In addition, the heating rate of the mode is around 80-150 phonon/s (drift from month to month), which is modeled as jump operators $\sqrt{\gamma_h}a$ and $\sqrt{\gamma_h}a^\dagger$. 

The Hamiltonians for the numerical simulation in QuTip are provided below. Motion heating is modeled using this master equation:
\begin{equation}
    \dot{\rho}=-i\left [ H, \rho \right ] + \gamma_{h} \mathcal{D}\left [ a^\dagger \right ] \rho + \gamma_{h} \mathcal{D}\left [ a \right ] \rho.
\label{Eq.Sim.Main}
\end{equation}

The Hamiltonian $H$ includes $H_{m}=\omega_z a^\dagger a$, displacement drive $H_{dis}=i\Omega a^\dagger e^{-i((\omega_z+\Delta) t + \phi_{dis})} + h.c.$ and spin-motion coupling Hamiltonians, which are listed below, with $e^{ikx}=e^{i\eta(a+a^\dagger)}$.

First-order blue sideband:
\begin{equation}
    H_{bsb}= 
\frac{\Omega_{bsb}}{2\eta} \left[ \sigma_+ e^{ikx} e^{-i (\delta_{bsb} t + \phi_{bsb})}+ h.c.\right],
\end{equation}
First-order red sideband:
\begin{equation}
H_{rsb}  = 
\frac{\Omega_{rsb}}{2\eta} \left[ \sigma_+ e^{ikx} e^{-i (\delta_{rsb} t + \phi_{rsb})}+ h.c.\right],
\end{equation}
Second-order red sideband:
\begin{equation}
H_{2rsb}  = 
\frac{\Omega_{2rsb}}{\eta^2} \left[ \sigma_+ e^{ikx} e^{-i (\delta_{2rsb} t + \phi_{2rsb})}+ h.c.\right], 
\end{equation}
Squeezing $H_{sq}=H_{sq,r_+}+H_{sq,r_-}+H_{sq,b_+}+H_{sq,b_-}$:
\begin{equation}
\begin{aligned}
H_{sq,r_\pm} &= \frac{ \Omega_{sq}} {2 \eta} \left\{ \sigma_+ e^{ikx} 
        e^{-i\left [(-\omega_z \pm \delta_m )t + \phi_{sq,r_\pm}\right]}  + h.c. \right\},\\
H_{sq,b_\pm} &= \frac{ \Omega_{sq}} {2 \eta} \left\{ \sigma_+ e^{ikx} 
        e^{-i\left [(\omega_z \pm \delta_m )t + \phi_{sq,b_\pm}\right]}  + h.c. \right\}.
\end{aligned}
\end{equation}

\section{Original data for dissipative enhanced synchronization}

The original data for Fig.~4(a) in the main text is shown in Fig.~\ref{Fig. S2}, together with the numerical simulations. The synchronization parameter $S$ of the deep quantum case is always smaller than the quantum regime, which is due to the larger quantum noise. The experimental data is slightly below the simulation line, which may due to the extra decoherence of motion, especially for the quantum regime cases.\\\\\\\\

%


\end{document}